\documentstyle[11pt,aaspp4,neu]{article}
\def\placefig#1{#1} \textheight 9.0in \topmargin -0.25in
\begin{document}

\def\dim#1{\mbox{\,#1}}
\def\fn{{f^{}_{\nu}}}

\title{Cosmological Neutrino Background Revisited}

\author{Nickolay Y.\ Gnedin}
\affil{Department of Astronomy, University of California, Berkeley, CA 94720;
       gnedin@astron.berkeley.edu}
\and
\author{Oleg Y.\ Gnedin}
\affil{Princeton University Observatory, Princeton, NJ 08544;
       ognedin@astro.princeton.edu}

\begin{abstract}
We solve the Boltzmann equation for cosmological neutrinos around the epoch
of the electron-positron annihilation in order to verify the freeze-out
approximation and to compute accurately the cosmological neutrino
distribution function.  We find the radiation energy density to be about
0.3\% higher than predicted by the freeze-out approximation.  As a result,
the spectrum of the Cosmic Microwave Background anisotropies changes by
$\sim 0.3-0.5\%$, depending on the angular scale, and the amplitude of the
mass fluctuations on scales below about 100 $h^{-1}\dim{Mpc}$ decreases by
about 0.2-0.3\%.
\end{abstract}

\section{Introduction}
Anisotropies in the Cosmic Microwave Background (CMB) provide a powerful
tool to probe the cosmological parameters (Hu, Sugiyama \& Silk 1997).  The
results of four-year work of the COBE satellite (Bennett et al.\ 1996) allow
us to determine the power spectrum of the CMB anisotropies to an accuracy of
7\%, but with relatively poor angular resolution, $\theta_{\rm FWHM} \approx
7^\circ$.  New planned satellite missions, MAP (Bennett et al. 1997) and
Planck (Bersanelli et al. 1997), will achieve an accuracy of better than 1\%
in the power spectrum with the sub-degree resolution (Bond, Efstathiou \&
Tegmark 1997; Zaldarriaga, Spergel \& Seljak 1997).  In turn, we need to
bring the theoretical models to the same level of accuracy.

In the standard cosmological model, after the epoch of the Big Bang
Nucleosynthesis the relativistic particles include photons and the three
species of neutrinos (Kolb \& Turner 1990; Peebles 1993).  While the
abundance of photons is directly measured from the CMB observations, the
abundance of primordial neutrinos can only be assessed theoretically.  The
standard way to perform such a calculation is to use the so called
freeze-out approximation, which assumes that neutrinos decouple
instantaneously from the rest of the universe at temperature of about
$4\dim{MeV}$.  Then the distribution function of all three neutrino species
retains the Fermi-Dirac form with the only parameter, the neutrino
temperature, uniquely tied to the observed CMB temperature.

However, even after decoupling the high-energy neutrinos still interact,
albeit slowly, with the electron-positron plasma, contrary to the basic
assumption of the freeze-out approximation.  This interaction leads to some
of the photon energy being transferred into neutrinos. But because it is the
photon energy density that is directly measured, the total energy density of
the universe in the relativistic species {\it relative\,} to the energy
density in photons (which is measured observationally to about 0.3\%
accuracy) will be somewhat higher than the one predicted by the freeze-out
approximation.

Several previous attempts have been made to compute cosmological neutrino
decoupling in greater detail, though still assuming that neutrino
distribution functions have Maxwellian form (Dicus et al.\ 1982; Herrera \&
Hacyan 1989; Raha \& Mitra 1991; Dolgov \& Fukugita 1992), with the most
comprehensive study given by Dodelson \& Turner (1992).  Recently, two more
papers have addressed this problem with the full account for the Fermi-Dirac
form of the neutrino distribution functions (Hannestad \& Madsen 1995;
Dolgov, Hansen, \& Semikoz 1997).  Both of these studies, however, have not
achieved the desired level of accuracy of the numerical calculation (about
$10^{-4}$, which is equivalent to a 1\% accuracy in a 1\% correction to the
freeze-out approximation).  The problem is complex: solution of the full
Boltzmann equation in three dimensions is at the very edge of modern
computing capabilities.  As a result, the previous calculations have only
been able to cover slightly more that two decades in the neutrino momentum,
which is insufficient to compute an asymptotic behavior of the neutrino
distribution function.

In this paper we complete the calculation of cosmological neutrino
decoupling using an extensive calculation on a parallel supercomputer,
placing special emphasis on achieving complete numerical convergence, and
covering over seven decades in the neutrino momentum.

The paper is composed in the following way.  We derive and solve the
Boltzmann equations for all three neutrino species in \S\ref{sec:BE}.  In
\S\ref{sec:NI}, we briefly touch upon the relevant numerical issues,
relegating the details of our numerical method to Appendix.  Finally in
\S\ref{sec:results}, we present our results for the neutrino energy density
and compare them to the freeze-out approximation.  We also obtain accurate
distribution functions for the cosmological neutrinos.

\section{Neutrino Kinetics in the Expanding Universe}
\label{sec:BE}

The Boltzmann equation for neutrinos in the expanding universe is
(Kolb \& Turner 1990):
\begin{equation}
	E_\nu {\partial \fn\over\partial t} - H q^2 {\partial \fn
	\over\partial E_\nu} =
	{\cal C}[\fn],
	\label{boleq}
\end{equation}
where $\fn(q,t)$ is the neutrino distribution function, $E_\nu$ is the
neutrino energy ($E_\nu = q$ since the neutrino mass, even if it exists, is
assumed to be much smaller than our characteristic energy scale, $\sim$
MeV), and ${\cal C}[\fn]$ is the collisional integral.  Hereafter we use
units in which $\hbar=c=1$.  In the case of neutrinos interacting with the
electron-positron pairs and other neutrino species via annihilation and
scattering reactions, the collisional integral is (Hannestad \& Madsen 1995)
\begin{equation}
	{\cal C}[\fn] = \sum
	{1\over2(2\pi)^5} \int 
	{d^3 p_2\over 2E_2}
	{d^3 p_3\over 2E_3}
	{d^3 p_4\over 2E_4}
	\Lambda(f_1,f_2,f_3,f_4) M^2\delta^4(\bar p_1+\bar p_2-\bar p_3-\bar p_4),
	\label{eq:collint}
\end{equation}
where $p_1\equiv q$, $f_1\equiv\fn$,
\[
	\Lambda(f_1,f_2,f_3,f_4) \equiv f_4f_3(1-f_2)(1-f_1) -
	f_1f_2(1-f_3)(1-f_4),
\]
$M^2$ is the matrix element squared and summed over initial and final spin
states, $\bar p_i$ are four-momenta of the incoming (1,2) and outgoing (3,4)
particles, and the sum is taken over all of the reactions involving $f_1$.

\def\tableone{
\begin{table}[b]
\caption{Neutrino Reactions\label{tabone}}
\medskip
$$
\begin{tabular}{lc}
Reaction & $M^2$ \\ \tableline
$\nu_i + \bar\nu_i \rightarrow e^- + e^+$ & 
$	32G_F^2\left[(C_V+C_A)^2Q_3+(C_V-C_A)^2Q_2+(C_V^2-C_A^2)Q_4\right]$ \\
$\nu_i + e^- \rightarrow \nu_i + e^-$ &
$	32G_F^2\left[(C_V+C_A)^2Q_1+(C_V-C_A)^2Q_3-(C_V^2-C_A^2)Q_5\right]$ \\
$\nu_i + e^+ \rightarrow \nu_i + e^+$ &
$	32G_F^2\left[(C_V+C_A)^2Q_3+(C_V-C_A)^2Q_1-(C_V^2-C_A^2)Q_5\right]$ \\
$\nu_i + \bar\nu_i \rightarrow \nu_i + \bar\nu_i$ & $128G_F^2Q_3$ \\
$\nu_i + \bar\nu_i \rightarrow \nu_j + \bar\nu_j$ & $ 32G_F^2Q_3$ \\
$\nu_i + \bar\nu_j \rightarrow \nu_i + \bar\nu_j$ & $ 32G_F^2Q_3$ \\
$\nu_i + \nu_k     \rightarrow \nu_i +     \nu_k$ & $ 32G_F^2Q_1$ \\
\end{tabular}
$$
\end{table}
}
\placefig{\tableone}

The list of all neutrino reactions is presented in Table \ref{tabone}, along
with the respective matrix elements (Hannestad \& Madsen 1995). Indecies
$i$, $j$, $k$ run over electron, muon, and tau neutrino, with the exception
that $j\not= i$.  The factor $G_F$ is the Fermi coupling constant, and
coefficients $C_V$ and $C_A$ for different types of neutrinos are given by
the following equations (for example, Kaminker et al.\ 1992):
\begin{eqnarray}
        C_V(\nu_e) & = & 2\sin^2\Theta_W+{1\over2},\;\;\;\;
        C_A(\nu_e) =  {1\over2}, \nonumber \\
        C_V(\nu_\mu,\nu_\tau) & = & 2\sin^2\Theta_W-{1\over2},\;\;\;\;
        C_A(\nu_\mu,\nu_\tau) = -{1\over2}, \nonumber
\end{eqnarray}
where $\Theta_W$ is the Weinberg angle, and we adopt $\sin^2\Theta_W=0.23$.

Quantities $Q_i$ are defined as follows:
\begin{eqnarray}
	Q_1 & = & (\bar p_1\cdot\bar p_2)(\bar p_3\cdot\bar p_4), \nonumber\\
	Q_2 & = & (\bar p_1\cdot\bar p_3)(\bar p_2\cdot\bar p_4), \nonumber\\
	Q_3 & = & (\bar p_1\cdot\bar p_4)(\bar p_2\cdot\bar p_3), \nonumber\\
	Q_4 & = & m^2 (\bar p_1\cdot\bar p_2), \nonumber\\
	Q_5 & = & m^2 (\bar p_1\cdot\bar p_3),
	\label{qfuncs}
\end{eqnarray}
where $m$ is the electron mass. As has been shown by Hannestad \& Madsen
(1995), integrals over $d^3p_4$ and over angles in $d^3p_2$ and $d^3p_3$ can
be computed analytically, yielding
\begin{equation}
	{\cal C}[\fn] = \sum
	{1\over2(2\pi)^5} \int 
	{p_2^2 dp_2\over 2E_2}
	{p_3^2 dp_3\over 2E_3}
	\Lambda(f_1,f_2,f_3,f_4) 
	F(p_1,p_2,p_3).
	\label{eq:collfin}
\end{equation}
In order to minimize the possibility of an error in the complicated factors
$F(p_1,p_2,p_3)$, we have used {\it Mathematica\/} software package to
perform the calculations.  The resultant expressions are too large to be
presented here, but the original {\it Mathematica\/} script and the FORTRAN
source code are available upon request.
	
In order to calculate the evolution of the neutrino distribution functions,
$\fn(\zeta)$, we need to include the equations describing the evolution of
the scale factor and the energy density of the universe:
\begin{equation}
        {da\over dt} = \left({8\pi G\over3}\rho\right)^{1/2}a,
        \label{adot}
\end{equation}
and
\begin{equation}
        {d\rho\over dt} = -3H(\rho+p),
        \label{rhodot}
\end{equation}
where $\rho$ and $p$ are the energy density and the pressure, respectively:
\[
        \rho(T) = {T^4\over \pi^2}\left[2C_{1/2}(\lambda)+{\pi^4\over 15}
        \right] + \rho_\nu,
\]
\[
        p(T) = {T^4\over \pi^2}\left[{2\over3}\left(C_{1/2}(\lambda)-
        \lambda^2C_{-1/2}(\lambda)\right)+{\pi^4\over 45}
        \right] + {\rho_\nu\over3},
\]
where $\lambda \equiv m/T$ and the functions $C_n(\lambda)$ are defined as
\[
        C_n(\lambda) \equiv \int_\lambda^\infty
        {\sqrt{x^2-\lambda^2}x^{2n+1}dx\over e^x+1}. 
\]
Here $\rho_\nu$ is the energy density in the three neutrino species,
\[
        \rho_\nu = {1\over\pi^2}\int_0^\infty q^3 
        \left[f_{\nu_e}(q)+f_{\nu_\mu}(q)+f_{\nu_\tau}(q)
        \right] dq.
\]
We also note that since the coefficients $C_V$ and $C_A$ are the same for
muon and tau neutrino, their distribution functions are equal.

\section{Numerical Issues}
\label{sec:NI}
 
Equation (\ref{boleq}), along with equations (\ref{adot}) and
(\ref{rhodot}), can now be integrated numerically for each of the neutrino
species.  In order to eliminate the derivative with respect to the neutrino
momentum in equation (\ref{boleq}), we employ the comoving momentum $\zeta =
qa$.  We lay out the neutrino distribution function on a logarithmically
spaced mesh in the range $10^{-5.5}\leq q/T \leq10^{1.7}$ with 40 points per
decade (289 points altogether).  By appropriately changing the limits and
sampling of the momentum mesh, we have verified that such a discretization
offers a fully convergent solution to an accuracy of better than $10^{-4}$.
After this procedure, equation (\ref{boleq}) becomes a system of coupled
ordinary differential equations.  We begin the integration at
$T=10\dim{MeV}$ and carry it out to $T=10^{-3}\dim{MeV}$, at which point the
desired precision is achieved.  The relative accuracy of the integration at
each time step is set to $10^{-7}$.

The resultant ordinary differential equations are stiff and present a
considerable computational challenge.  Standard methods require computing
the full Jacobian, which is virtually impossible for our fairly complicated
system of equations.  For the purpose of this calculation, we develop a
special numerical scheme, presented in the Appendix, which can handle stiff
equations of the Bolztmann type and does not require computing the full
Jacobian.  Our scheme is more efficient by a factor of 20 to 60 than
the standard fifth order adaptive Runge-Kutta method.

Finally, we note that the most time consuming part, calculating the
collisional integral, can be done very efficiently in parallel. Our final
computation has been performed on the NCSA Power Challenge Array with 12
R10000 processors and has consumed about 200 processor-hours.

\section{Results and Discussion}
\label{sec:results}

\def\tableonea{
\begin{table}
\caption{Main Results\label{tabonea}}
\medskip
$$
\begin{tabular}{lcc}
Quantity & Exact value & Freeze-out approximation \\ \tableline
$(aT)_{\rm before}/(aT)_{\rm after}$            & 0.7144 & 0.7138 \\
$\epsilon_R/(a_RT_{0,\gamma}^4)$                & 1.6863 & 1.6813 \\
$\epsilon_{\nu_e}/(a_RT_{0,\gamma}^4)$          & 0.2293 & 0.2271 \\
$\epsilon_{\nu_\mu}/(a_RT_{0,\gamma}^4)$        & 0.2285 & 0.2271 \\
$\epsilon_{\nu_\tau}/(a_RT_{0,\gamma}^4)$       & 0.2285 & 0.2271 \\
$n_{\nu_e}/n_\gamma$                            & 0.2745 & 0.2727 \\
$n_{\nu_\mu}/n_\gamma$                          & 0.2739 & 0.2727 \\
$n_{\nu_\tau}/n_\gamma$                         & 0.2739 & 0.2727 \\
\end{tabular}
$$
\end{table}
}
\placefig{\tableonea}

The main outcome of our calculations is the number density and the energy
density of all three neutrino species at the current epoch relative to those
of photons.  The results are presented in Table \ref{tabonea}. For
comparison, we also give the respective numbers computed in the freeze-out
approximation.  All values are accurate to the last decimal place shown.
The most important quantity, the total radiation energy density in the
universe, differs from the freeze-out approximation by only 0.3\%.

Another way of presenting this difference is the effective number of neutrino
species, $N_{\rm eff}$. We can rewrite
the expression for the energy density of the universe as
\begin{equation}
	\epsilon_R = \left[ 1 + N_{\rm eff} {7\over8}
	\left(4\over11\right)^{4/3}\right]
	a_R T_{0,\gamma}^4,
	\label{eq:raden}
\end{equation}
where in the freeze-out approximation $N_{\rm eff}=3$.  
From Table \ref{tabonea} we obtain:
\[
	N_{\rm eff} = 3.022.
\]
This number does not, of course, mean that there are more than 3 species of
neutrinos; it is simply a number that should be used in the freeze-out
approximation to reproduce the exact result.  Since most of the previously
obtained results and existing numerical codes are based on the freeze-out
approximation, it is convenient to use $N_{\rm eff}$: one simply has to use
3.022 instead of 3.0 in every place in the code where the neutrino energy
density is computed.

We note here that we find a somewhat larger effect than both Hannestad \&
Madsen (1995) and Dolgov et al.\ (1997), who found $N_{\rm eff}=3.017$ and
$N_{\rm eff}$ from 3.013 to 3.019 (depending on the method of calculation)
respectively.  We attribute this difference to the higher accuracy of our
calculation and the lack of numerical convergence in the previous work. In
particular, when we adopt a momentum range from $q/T=10^{-1}$ to $10^{1.3}$,
as in Dolgov et al. (1997), we obtain $N_{\rm eff}=3.019$, in agreement with
the authors.  If we further reduce the momentum range from $q/T = 10^{-0.3}$
to $10^{1.1}$, as in Hannestad \& Madsen (1995), we recover their result,
$N_{\rm eff}=3.017$.

\def\capTE{ The fractional difference between the effective neutrino
temperature and its value in the freeze-out approximation, as a function of
the neutrino momentum, $q$.  Solid line is for the electron neutrino
temperature, and dashed line is for the muon and tau neutrino temperatures.}

\placefig{
\begin{figure}
\plotone{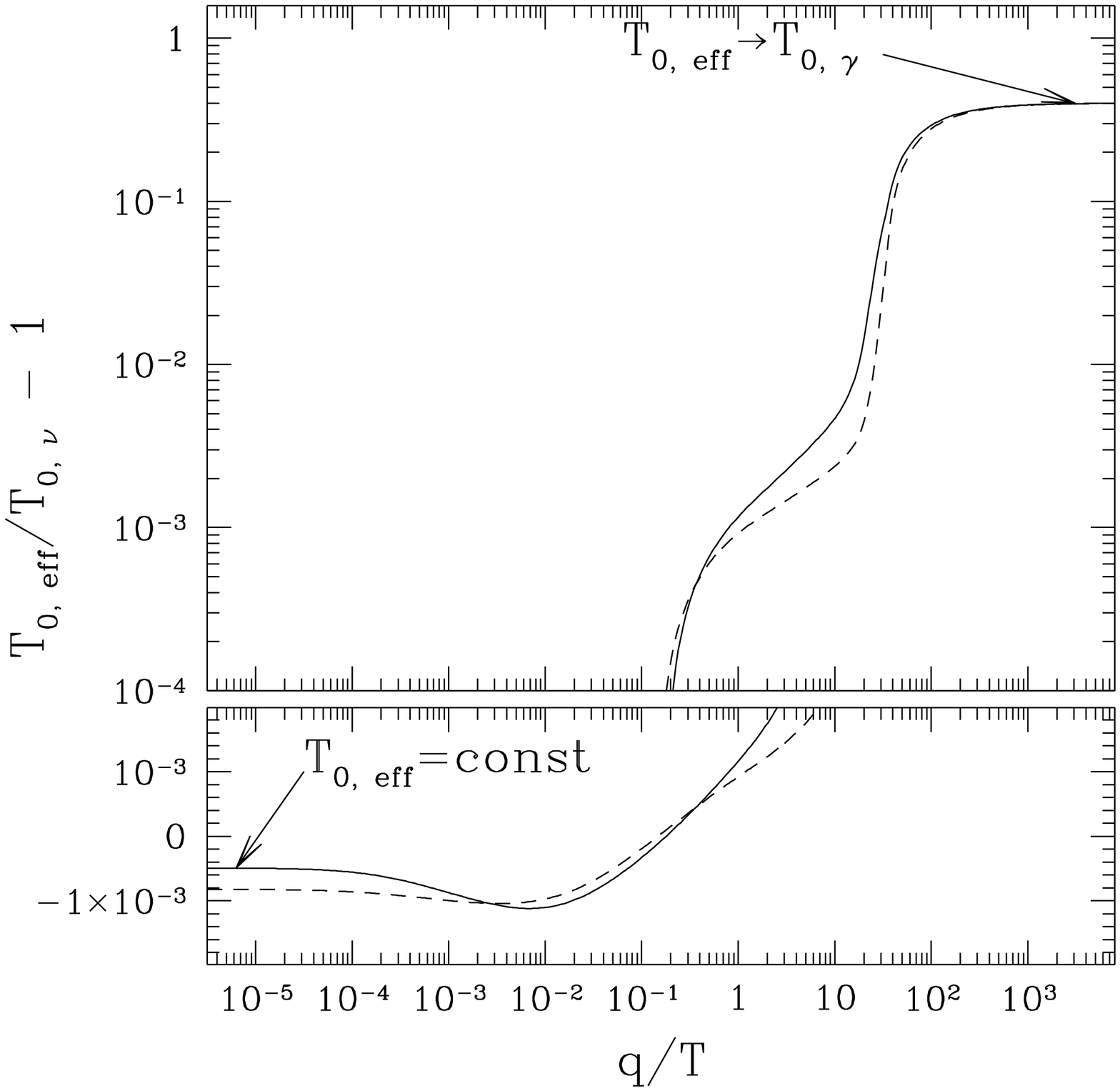}
\caption{\capTE\label{figTE}}
\end{figure}
}

We can also characterize the final neutrino distribution function. Let us
introduce the effective neutrino temperature, $T_{\rm eff}$, as
\begin{equation}
	\fn(q) \equiv {1\over e^{q/T_{\rm eff}}+1}.
\end{equation}
Since the neutrino distribution function is not of the Fermi-Dirac form any
more, $T_{\rm eff}$ is a function of the neutrino momentum $q$.  Figure 1
shows the deviation of the effective temperature at the current epoch, $T_{0,
\rm eff}$, from the freeze-out approximation value, 
$T_{0,\nu}\equiv(4/11)^{1/3} T_{0,\gamma}$,
as a function of $q/T$ (this ratio is independent of time 
after
electron-positron annihilation).  For the high neutrino momenta, the effective
neutrino temperature asymptotically approaches the photon temperature,
because the high-energy neutrinos can efficiently interact with the electrons
via pair creation even after annihilation,
\[
	T_{\rm eff}\rightarrow T\ \ \ {\rm for}\ q \rightarrow \infty.
\]
For the low momenta, $q \ll T$, neutrino interactions with electrons and
positrons are suppressed by a factor $q^2$ and do not affect the evolution
of the distribution function. However, the rate of the neutrino-neutrino
interactions is proportional to the first power of the momentum $q$. Thus,
in the limit of small momenta,
\[
	{\partial \fn(q)\over\partial t} = 
	{q\over T_{\rm eff}^2} {d T_{\rm eff}\over dt} 
	\propto q,
\]
and we find that
\[
	T_{\rm eff}\rightarrow {\rm const}\ \ \ {\rm for}\ q \rightarrow 0.
\]
The value of the constant is 0.99950 $T_{0,\nu}$ for the electron neutrinos
and 0.99918 $T_{0,\nu}$ for the muon and tau neutrinos.

Our results have several immediate cosmological implications.  First, the
change of the radiation energy density of the universe will affect the
spectrum of CMB anisotropies at about 0.3\% level just behind the first
acoustic peak, $l\sim 300$, and at about 0.5\% level at the damping scale,
$l\sim1000$ (Hu et al.\ 1995).  If more than 0.5\% accuracy is required in
calculating the CMB anisotropies, equation (\ref{eq:raden}) should be used.

Second, the change of the radiation energy density of the universe affects
the evolution of linear density fluctuations on galactic and subgalactic
scales.  We have computed the matter transfer function using {\it COSMICS\/}
package (Bertschinger 1995) and we find that for a cosmological model with
$\Omega_0=1$, $h=0.5$, and $\Omega_b=0.05$, the rms density fluctuation at
$8h^{-1}\dim{Mpc}$ scale, $\sigma_8$, decreases by about 0.2\%, and the rms
density fluctuation on $100h^{-1}\dim{kpc}$ scale decreases by about 0.3\%.
These latter changes, however, are too small to be of any interest in the
foreseeable future.

In addition, neutrino decoupling affects primordial helium
production. However, Dodelson \& Turner (1992) showed that the net change in
the primordial helium abundance is virtually unobservable due to
cancellation of two competing effects: one is the higher expansion rate,
which leads to a higher helium abundance, and the other is the faster
neutron decay rate, which leads to a lower helium abundance.  A small change
in the neutrino number density will produce a small change in the massive
neutrino mass density (Hannestad \& Madsen 1995), but this change is again
too small to be of any practical interest.

Thus, we conclude that only in computing the CMB anisotropies should one
need to worry about the accurate calculation of neutrino decoupling;
otherwise, the effect is negligibly small.  Overall, our calculations
confirm the validity of the freeze-out approximation. If the accuracy of a
few percent is sufficient, one can safely use the freeze-out approximation
to compute any property of cosmological neutrinos.

\acknowledgements
We are grateful to D.\ Yakovlev, J.\ Ostriker, D.\ Spergel, 
J.\ Madsen, B.\ Fields,
S.\ Hannestad, and A.\ Dolgov for valuable comments.
We thank the anonymous referee for pointing us to an error in the
original manuscript. 
N.\ G.\ was supported by the UC Berkeley grant 1-443839-07427.
Calculations were performed on the NCSA Power Challenge Array under the
grant AST-960015N.

\appendix

\section{Numerical Method}

In this paper we are dealing with a particular kind of an ordinary differential
equation that can be presented in the following form:
\begin{equation}
	{dy\over dt} = f(y) \equiv w(y) - k(y)y,
	\label{difeq}
\end{equation}
where both $w$ and $k$ are slow functions of $y$, but not of $t$,
and $k$ is positive.
This equation is stiff, and a numerical method which does not handle
stiff equations requires a time step $\Delta t$ such that
\[
	k \Delta t\ll 1.
\]
Numerical methods that can deal with stiff equations usually have a much
less stringent restriction on the time step,
\[
	\left|\partial w\over\partial y\right|\Delta t  \ll 1,
\]
and, because $w$ is a slow function of $y$, we assume that
\[
	\left|\partial w\over\partial y\right| \ll k.
\]
However, standard techniques for stiff equations require computing the
full Jacobian,
\[
	J = {\partial w\over\partial y} - {\partial k\over\partial y} y - k.
\] 
In our case this quantity is very difficult to
compute, because $w$ is an integral over $y$ and numerical evaluation of the
integral involves nontrivial interpolation.

We therefore proceed differently and design a numerical scheme which
involves only a partial Jacobian,
\[	
	\hat{J} \equiv -k,
\]
which can be computed simultaneously with the r.h.s.\ of equation (\ref{difeq})
at no extra cost. 

Additional advantage of using the partial Jacobian $\hat{J}$ instead
of the full Jacobian $J$ is that for a system of equations, the partial 
Jacobian is a diagonal matrix, which can be inverted much faster than
the full Jacobian, which is usually a general matrix.

However, we cannot simply take a standard numerical scheme and replace
the full Jacobian with the partial one, because the different orders
of the numerical error will not cancel out in this case. Thus, we need
to design a special scheme which will assure the proper cancellation
of the numerical error up to a given order.

The numerical scheme to update $y$ from $y=y_0$ to $y=y_1$ in a time
interval $h$ is constructed as follows:
\begin{eqnarray}
	g_1 & = & {h f(y_0)\over1+\gamma h k},\nonumber\\
	g_2 & = & {h f(y_0 + a_{21} g_1) + c_{21} g_1 \over 1+\gamma h k},
	\nonumber\\
	g_3 & = & {h f(y_0 + a_{31} g_1 + a_{32} g_2) + c_{31} g_1 + c_{32} g_2
	\over1+\gamma h k},\nonumber\\
	y_1 & = & y_0 + b_1 g_1 + b_2 g_2 + b_3 g_3,
	\label{numscheme}
\end{eqnarray}
where $\gamma$, $a_i$, $b_i$, and $c_i$ are constants. The values for
$\gamma$, $a_{21}$, $a_{31}$, $a_{32}$, and $c_{21}$ are given in Table
\ref{tabtwo}.

\def\tabletwo{
\begin{table}
\caption{\label{tabtwo}}
\medskip
$$
\begin{tabular}{cc}
Constant & Value \\ \tableline
$\gamma$ & 0.788675134594812882251 \\
$a_{21}$ & 1\\
$a_{31}$ & 0.56698729810778067662\\
$a_{32}$ & 1/4\\
$c_{21}$ & -1.26794919243112270647\\
\end{tabular}
$$
\end{table}
}
\placefig{\tabletwo}

By varying the remaining constants, we construct two numerical schemes: the
third order and the second order, respectively (Table \ref{tabthr}). Thus,
the difference between the values of $y_1$ computed with the two schemes can
serve as an estimate of numerical errors.  Again, {\it Mathematica\/} was
used to compute the values of the constants that give the cancellation of
numerical errors to the required order.

\def\tablethr{
\begin{table}
\caption{\label{tabthr}}
\medskip
$$
\begin{tabular}{ccc}
Constant & Third order scheme & Second order scheme \\ \tableline
$c_{31}$ & -3/2                    & -3.183012701892219323 \\
$c_{32}$ & -1.1830127018922193234  & -3.049038105676658006 \\
$b_1$    &  1.3779915320718537844  &  1.566987298107780677 \\
$b_2$    &  0.9553418012614795489  &  1.038675134594812883 \\
$b_3$    &  2/3                    &  0.21132486540518711775 \\
\end{tabular}
$$
\end{table}
}
\placefig{\tablethr}

\placefig{\end{document}}

\clearpage
 
\tableone
 
\clearpage
 
\tableonea
 
\clearpage
 
\tabletwo
 
\clearpage
 
\tablethr
 
\clearpage

\newcounter{figurecap}
\setcounter{figurecap}{0}
 
\begin{center}
\bf Figure Captions
\end{center}
 
\refstepcounter{figurecap}
Fig.\ \thefigurecap---\label{figTE}\capTE

\end{document}